\documentclass[%
 aip,
 amsmath,amssymb,
 reprint%
]{revtex4-1}

\usepackage{graphicx}
\usepackage{dcolumn}
\usepackage{bm}

\usepackage[normalem]{ulem} 

\usepackage{array}
\newcolumntype{K}{>{\centering\arraybackslash}b{0.92cm}}

\usepackage[utf8]{inputenc}
\usepackage[T1]{fontenc}
\usepackage{mathptmx} 
\usepackage{booktabs}
\usepackage{multirow}
\usepackage[version=4]{mhchem}
\usepackage{amsmath}
\usepackage{xcolor}
\definecolor{aaltoOrange}{RGB}{255,121,0}%
\definecolor{aaltoBlue}{RGB}{0,101,189}%

\DeclareSymbolFont{epsilon}{OML}{ntxmi}{m}{it}
\DeclareMathSymbol{\epsilon}{\mathord}{epsilon}{"0F}
\mathchardef\mhyphen="2D
\usepackage{physics}

\begin{document}

\preprint{AIP/123-QED}

\title[Relativistic correction scheme for core-level binding energies from $GW$]{Relativistic correction scheme for core-level binding energies from $GW$}

\author{Levi Keller}
 \email{Levi.Keller@aalto.fi.}
\affiliation{Department of Applied Physics, Aalto University, Otakaari 1, FI-02150 Espoo, Finland}%

\author {Volker Blum}

\affiliation{\small{Department of Mechanical Engineering and Materials Science, Duke University, Durham, North Carolina 27708, USA}}

\author{Patrick Rinke}
\affiliation{Department of Applied Physics, Aalto University, Otakaari 1, FI-02150 Espoo, Finland}%

\author{Dorothea Golze}%
\affiliation{Department of Applied Physics, Aalto University, Otakaari 1, FI-02150 Espoo, Finland}%

\date{\today}

\begin{abstract}

We present a relativistic correction scheme to improve the accuracy of 1s core-level binding energies calculated from Green's function theory in the $GW$ approximation, which does not add computational overhead. An element-specific corrective term is derived as the difference between the 1s eigenvalues obtained from the self-consistent solutions to the non- or scalar-relativistic Kohn-Sham equations and the four-component Dirac-Kohn-Sham equations for a free neutral atom. We examine the dependence of this corrective term on the molecular environment and on the amount of exact exchange in hybrid exchange-correlation functionals. This corrective term is then added as a perturbation to the quasiparticle energies from partially self-consistent and single-shot $GW$ calculations. We show that this element-specific relativistic correction, when applied to a previously reported benchmark set of 65 core-state excitations [J. Phys. Chem. Lett. 11, 1840 (2020)], reduces the mean absolute error (MAE) with respect to experiment from 0.55 to 0.30~eV and eliminates the species dependence of the MAE, which otherwise increases with the atomic number. The relativistic corrections also reduce the species dependence for the optimal amount of exact exchange in the hybrid functional used as starting point for the single-shot $G_0W_0$ calculations. Our correction scheme can be transferred to other methods, which we demonstrate for the Delta self-consistent field ($\Delta$SCF) approach based on density functional theory.
\end{abstract}

\maketitle

\section{Introduction}
Core-level binding energies (BEs), measured by X-ray photoemission spectroscopy (XPS), are element-specific, but depend also on the local chemical environment and thus afford access to information about the chemical bonding, oxidation state, and coordination of a given element in a sample.\cite{Bagus2013,Bagus1999,Siegbahn1969all} The energetic differences (chemical shifts) between atomic species of the same type can be smaller than 0.5~eV for second-row elements and can be as low as 0.1~eV for carbon~1s excitations.\cite{Pireaux1976} The interpretation of an XPS spectrum can be very difficult due to overlapping features or the lack of well-defined reference data.\cite{Aarva2019a,Aarva2019b} Highly accurate theoretical tools for the prediction of relative and absolute BEs are therefore necessary to guide the experiment and its interpretation. The reliable computation of absolute core-level energies is generally more challenging than the calculation of energy shifts\cite{Vines2018} and is the focus of this work. \par
The most common approach to calculating core-level BEs is the Delta self-consistent field ($\Delta$SCF) method,\cite{Bagus1965} wherein one computes the total energy difference between the ground and core-ionized state using Kohn-Sham density functional theory (KS-DFT). The best absolute core-level BEs have been obtained with meta-generalized gradient approximation (meta-GGA) functionals, yielding mean deviations of $\approx$0.2~eV with respect to experiment for small molecules.\cite{Bellafont2016b,Kahk2019} Similar accuracy has been obtained with high-level wavefunction-based Delta coupled cluster  methods,\cite{Zheng2019,Holme2011,Sen2018} albeit at much higher computational cost. The introduction of occupation constraints and the explicit generation of a charged system in $\Delta$-based approaches leads to a plethora of problems.\cite{Michelitsch2019,Bagus2019} Most importantly, the application to periodic systems requires further approximations, e.g., neutralizing the unit or supercell.\cite{Bellafont2017,Koehler2004}\par 
These problems do not occur in response theories, which avoid the explicit introduction of a core-ionized system and recently emerged as viable alternatives to $\Delta$-based schemes for core-level calculations. \cite{Liu2019,Voora2019,Golze2020} One particularly promising response approach is the $GW$ approximation\cite{Hedin1965} to many-body perturbation theory. $GW$ offers access to the core-state quasiparticle energies, which can be directly related to the core-level BEs. The $GW$ approach is the standard approach to compute valence excitations (band structures) in solid-state physics.
\cite{Golze2019} With the advent of efficient implementations with localized basis-sets,\cite{Blase2011, Ren2012, vanSetten2013, Bruneval2016, Wilhelm2016,Wilhelm2017} $GW$ has also become the method of choice for calculating the BEs of frontier orbitals in molecules and is nowadays routinely applied to large molecular structures with several hundred atoms.\cite{Wilhelm2016,Wilhelm2018,Wilhelm2019,Stuke2020} $GW$ has also been successfully applied to deep valence and semi-core states with excitation energies up to 80~eV.~\cite{Guzzo2011,Zhou2015,Kas2016}  However, the application of $GW$ for deep core states with BEs larger than 100~eV, which are the ones relevant for chemical analysis, has rarely been attempted. The first $GW$ studies\cite{Ishii2010,Aoki2018,Setten2018,Voora2019} for deep core states reported partly large deviations of several electronvolts from experiment. Recently, we have shown that $GW$ can also be successfully applied for 1s molecular core states when utilizing  highly accurate techniques for the integration of the self-energy\cite{Golze2018} and eigenvalue self-consistency in the Green's function.\cite{Golze2020}\par
The application of $GW$ (or any other method) to absolute core-levels binding energies requires an accurate treatment of relativistic effects. Already for the 1s core levels of the $p$-block 2nd period elements, the magnitude of relativistic effects begins to exceed the size of the chemical shifts. Recently, we have applied the $GW$ method to a benchmark set of 65 core-level BEs of the second-period elements C, N, O and F in small and medium-sized molecules (henceforth CORE65 benchmark set), obtaining a mean deviation from experiment of 0.3~eV.\cite{Golze2020} Therein, we applied a simple relativistic correction to the $GW$ computed core-level energies. The purpose of the present article is to describe this correction.\par 
Relativistic effects enter the $GW$ formalism via the underlying reference calculation. A fully-relativistic one-particle reference is described by a four-component Dirac spinor, or approximately by two-component spinors. Both types of spinors can describe noncollinear electronic states and thus spin-dependent electron-electron interactions. Explicitly spin-dependent $GW$ equations using spinors as input were only developed 12 years ago,\cite{Aryasetiawan2008,Aryasetiawan2009} and provide a framework to properly describe spin-orbit coupling (SOC) effects in $GW$. Several $GW$ codes emerged over the past years that implement these equations, employing spinors from all-electron 2-component Dirac-KS-DFT calculations,\cite{Kuehn2015,Holzer2019} from KS-DFT with second-variation SOC\cite{Sakuma2011} or from KS-DFT with fully-relativistic pseudopotentials.\cite{Scherpelz2016,Schuler2019} These implementations are in the following referred to as fully-relativistic $GW$ approaches.\par
Fully-relativistic $GW$ calculation have been primarily used to compute valence and conduction bands of solids with strong SOC effects. Fully-relativistic results have been reported for actinide metals,\cite{Kutepov2012,Ahmed2014} transition metal chalcogenides\cite{Sakuma2011,Scherpelz2016} and dichalcogenides,\cite{Molina-Sanchez2013,Schuler2019,Guo2019} perovskites\cite{Umari2014,Scherpelz2016} as well as bismuth-based topological insulators.\cite{Nechaev2013,Aguilera2013,Aguilera2013b,Aguilera2015,Nechaev2015,Battiato2017} Compared to non-relativistic $GW$, the fully-relativistic approach is computationally at least four times more expensive.\cite{Holzer2019} Alternatively, the SOC can be added as a perturbative correction to the quasiparticle band structure, which was common in earlier $GW$ studies.\cite{Rohlfing1998} Since this approach is computationally less expensive, it continues to be employed.\cite{Gao2018,Bushick2019,Zhang2019} However, it has been shown that the SOC post-correction scheme can fail, for example, to describe the band inversion in topological insulators.\cite{Aguilera2013b}\par
An explicit treatment of the SOC is typically not necessary for valence excitations of molecules, unless very heavy atoms are involved. Since molecular $GW$ studies have mainly focused on organic semiconductors,\cite{Golze2019} fully-relativistic $GW$ calculations are rare and have been mostly conducted for diatomic molecules,\cite{Kuehn2015,Scherpelz2016} but recently also for transition metal halide complexes.\cite{Holzer2019} If relativistic effects are considered in molecular $GW$ calculations at all, they are more commonly included by employing one-component reference states that capture only scalar-relativistic effects. This approach has been employed in $GW$ calculations with scalar-relativistic pseudopotentials\cite{Wilhelm2016,Blase2011} or the zeroth-order regular approximation (ZORA).\cite{Stuke2020}\par
Relativistic considerations for the innermost core-levels differ from those for valence excitations. SOC leads, for core states with angular quantum number $l>0$, to a splitting. Already for $2p$ states of third-row elements, e.g. sulfur, this splitting lies in the range of 1~eV and increases for 4th-period elements to several tens of eV.\cite{xraydata} SOC affected core states require thus a noncollinear treatment and we are only aware of one $GW$ study\cite{Kuehn2015} that reports results for deep, spin-orbit-coupled $p$ states.
In this study we focus on organic molecules and small inorganic molecules with elements C, N, O and F, for which typically the 1$s$ excitations in the energy range from 250 to 700~eV are measured in XPS. While scalar relativistic effects are heightened in the proximity of the poorly-screened nuclear charge, the SOC operator does not directly affect core states. In contrast to valence excitations, the most common scalar relativistic approximation, ZORA, performs poorly for the absolute eigenvalue associated with the innermost core levels.\cite{vanLeeuwen1994, Sundholm2002} In order to avoid the computational expense of a fully Dirac-like, two-component $GW$ calculation, which is not essential to the physics of 1s core excitations, we derive here an element-specific relativistic corrective term for non-relativistic and scalar-relativistic reference states, which we add in a post-$GW$ perturbative step.\par 
The remainder of this article describes this correction. In section \ref{sec:theory} we describe the $GW$ formalism, highlighting aspects that are particularly relevant for core-level calculations, and follow this up with an overview of the aspects of relativistic theory relevant to this work. We then describe the methods employed in section \ref{sec:methods}. We present and discuss the results of our correction schemes, which we apply to $GW$ and $\Delta$SCF computed core-level BEs of the CORE65 benchmark set in section \ref{sec:results} and finally draw conclusion in section~\ref{sec:conclusion}.

\section{\label{sec:theory} Theory}

\subsection{$\boldsymbol{G_0W_0}$ quasiparticle energies}
In practice, $GW$ is often performed as one-shot perturbative approach ($G_0W_0$) on top of an underlying mean field theory calculation. Possible mean field theories are Hartree-Fock (HF), KS-DFT or hybrid DFT which yield the molecular orbitals (MOs) $\{\phi_n\}$ and eigenvalues $\{\epsilon_n\}$ used as input (starting point) for the $G_0W_0$ calculation.
The $G_0W_0$ quasiparticle (QP) energies $\epsilon_n^{G_0W_0}$ are obtained by solving the QP equation 
\begin{equation}
\epsilon_n^{G_0W_0} = \epsilon_n + \mathrm{Re}\Sigma_n(\epsilon_n^{G_0W_0}) - v_n^{\mathrm{XC}}
\label{eq:epsilon_n}
\end{equation}
and can be related to the BE of state $n$ by $\mathrm{BE}_n=-\epsilon_n^{G_0W_0}$. The $n$th diagonal elements of the KS exchange-correlation (XC) potential and self-energy operator $\Sigma$ are denoted by 
 $v_n^{\mathrm{XC}} = \bra{\phi_n}v^{XC}\ket{\phi_n}$ and $\Sigma_n = \bra{\phi_n}\Sigma\ket{\phi_n}$, respectively. Note that we have omitted spin indices.
The self-energy operator is given by
\begin{equation}
 \Sigma(\bm{r},\bm{r}',\omega) = \frac{i}{2\pi}\int \mathrm{d}\omega' G_0(\bm{r},\bm{r}',\omega+\omega')W_0(\bm{r},\bm{r}',\omega')e^{i\omega'\eta}
 \label{eq:Sigma}
\end{equation}
where $G_0$ is the non-interacting KS Green's function, $W_0$ is the screened Coulomb interaction, and $\eta$ is a positive infinitesimal. The KS Green's function is obtained from the KS orbitals and eigenvalues by
\begin{equation}
 G_0(\bm{r},\bm{r}',\omega) = \sum_m\frac{\phi_m(\bm{r})\phi_m(\bm{r'})}{\omega - \epsilon_m - i\eta \mathrm{sgn}(\epsilon_F - \epsilon_m)}
 \label{eq:G0}
\end{equation}
where $\epsilon_\mathrm{F}$ is the Fermi energy, and the sum runs over both occupied and virtual KS orbitals. Within the random phase approximation, the screened Coulomb interaction is given by
\begin{equation}
 W_0(\bm{r},\bm{r}',\omega) = \int \mathrm{d}\bm{r}'' \frac{\varepsilon^{-1}(\bm{r},\bm{r}'',\omega)}{|\bm{r}''-\bm{r}'|}
\end{equation}
where the dielectric function $\varepsilon$ is 
\begin{equation}
 \varepsilon(\bm{r},\bm{r}',\omega) = \delta(\bm{r},\bm{r}') - \int \mathrm{d}\bm{r}'' \frac{\chi_0(\bm{r}'',\bm{r}',\omega)}{|\bm{r}-\bm{r}''|}
\end{equation}
and the irreducible polarizability $\chi_0$ is given in the real-space Adler-Wiser representation\cite{Adler1962,Wiser1963} as
\begin{align}
 \chi_0(\bm{r},\bm{r}',\omega) &= \sum_i^{\rm occ} \sum_a^{\rm virt} \phi_a(\bm{r}')\phi_i(\bm{r}')\phi_a(\bm{r})\phi_i(\bm{r}) \nonumber\\ &\times\bigg[\frac{1}{\omega - (\epsilon_a - \epsilon_i) + i\eta} + \frac{1}{-\omega - (\epsilon_a - \epsilon_i) + i\eta}\bigg]
\end{align}
The index $i$ runs over occupied orbitals, and $a$ runs over virtual orbitals.\par
Equation~\eqref{eq:epsilon_n} is non-linear, and can be solved iteratively or approximately by linearization using a Taylor expansion to first order around $\epsilon_n$.\cite{Golze2019} As we pointed out in our previous work,\cite{Golze2018,Golze2020} the linearization error increases rapidly with increasing BE and can already amount to 0.5~eV for deeper valence states.\cite{Golze2019} The magnitude of this error is in the range of the chemical shifts expected for 1s excitations. In addition, core-level BEs are an order of magnitude larger than deep valence BEs, potentially leading to even larger linearization errors. We therefore always solve the QP equation iteratively.

\subsection{Frequency treatment for core states}
The accurate frequency integration of the self-energy (Equation~\eqref{eq:Sigma}) is one of the major challenges for the calculation of deep core states.
A common approach for valence states is to evaluate the self-energy for imaginary frequencies and analytically continue it to the real frequency axis by fitting the self-energy matrix elements to a multipole model. Analytic continuation is employed in many state-of-the-art $GW$ implementations\cite{Liu2016,Wilhelm2016,Wilhelm2018,Ren2012} and yields accurate results for valence states.\cite{Setten2015} 
The structure of $\Sigma_n$ for a valence state is typically smooth in the frequency region where the QP solution is expected, and is well reproduced by analytic continuation. For core states, the self-energy has a complicated structure with many poles. We showed that analytic continuation becomes numerically unstable in the core region and completely fails to reproduces the self-energy structure.\cite{Golze2018} \par
For core states, a more accurate evaluation of the self-energy on the real frequency axis is required. We employ the contour deformation (CD) technique,\cite{Gonze2009,Blase2011,Govoni2015,Golze2018,Golze2019} where the numerically unstable integration along the real frequency axis is avoided by extending the integrand to the complex plane. The contours are chosen such that only the poles of $G_0$ are enclosed in the contours and the contour integral is evaluated using the residue theorem; see Refs.\citenum{Golze2018} and \citenum{Golze2019} for details. The integral along the real frequency axis is then evaluated as
\begin{equation}
 \begin{split}
  \Sigma(\bm{r},\bm{r}',\omega)& ={}  -\frac{1}{2\pi} \int_{-\infty}^{\infty} d\omega' G_0(\bm{r},\bm{r}',\omega+i\omega') W_0(\mathbf{r},\mathbf{r}',i\omega') \\
                   -&\sum_i\phi_i(\bm{r})\phi_i(\bm{r}')W_0(\bm{r},\bm{r}',|\epsilon_i-\omega|+i\eta)\theta(\epsilon_i-\omega)\\
                 + &\sum_a\phi_a(\bm{r})\phi_a(\bm{r}')W_0(\bm{r},\bm{r}',|\epsilon_a-\omega|+i\eta)\theta(\omega-\epsilon_a)
 \end{split}
 \label{eq:cd}
\end{equation}
where $\theta$ is the Heaviside step function and $i$ refers to occupied and $a$ to unoccupied orbitals. \par
The CD technique reproduces the self-energy structure for core excitations exactly and matches the results from the computationally more expensive fully-analytic solution of Equation~\eqref{eq:Sigma}.\cite{Golze2018} Recently, combining the CD approach with analytic continuation of $W_0$ has been proposed as alternative approach.\cite{Duchemin2020}

\subsection{Restoration of the core-level quasiparticle peak}
In this section we briefly present the $GW$ variants that we used in this work. By now many different $GW$ flavours have emerged in practical calculations. \cite{Golze2019} The most common flavor, $G_0W_0$ based on a semi-local DFT starting point, breaks down for core states, as we will detail in the following. We therefore need to go beyond the most common approach.

The problem with the conventional $GW$ approach is related to a loss of spectral weight in the QP peak in the spectral function $A(\omega)=1/\pi\sum_m\mathrm{Im}G_m$, where $m$ runs over all occupied and virtual states and $G=G_0+G_0\Sigma G$. For molecular valence states, $G_0W_0$ performed on top of a DFT calculation with the Perdew-Burke-Ernzerhof 
(PBE)\cite{Perdew1996a} functional ($G_0W_0@$PBE) yields a clearly identifiable QP peak. This peak corresponds to a distinct solution of Equation (\ref{eq:epsilon_n}). Multiple solutions, that would indicate spectral weight transfer to other peaks in the spectral function, have only been observed for a few systems for frontier orbitals.\cite{Lischner2012,Setten2015,Loos2018,Veril2018,Setten2018} These are rare cases and usually not more than two possible solutions are observed.\par 
The situation is dramatically different for deep core states. The analysis of the spectral functions in our recent work showed that a unique QP solution is not obtained with $G_0W_0@$PBE for 1s states.\cite{Golze2020} The spectral function shows a multitude of peaks with similar spectral weight, but no distinct QP excitation. A spurious transfer of spectral weight from the QP peak to plasmon satellites has been previously observed for deep valence states of transition metal oxides\cite{Gatti2015,Byun2019} and semi-core
excitations of sodium.\cite{Zhou2015} However, for deep core states the transfer of spectral weight is far more extreme resulting in a spectral function, where satellite spectrum and quasiparticle have completely merged. Such a spectral function contradicts the expected physics.
Photoemission spectra of molecular 1s excitations show strong QP peaks accompanied by satellites features due to multi-electron excitations such as shake-up processes, which are orders of magnitudes smaller than the main excitation.\cite{Sankari2006,Schirmer1987}\par
We showed that the 1s QP peak can be correctly restored by including eigenvalue self-consistency in $G$, while keeping PBE as starting point.\cite{Golze2020} This scheme is referred to as ev$GW_0@$PBE. In ev$GW_0$, the screened Coulomb interaction is kept fixed at the $W_0$ level and the Green's function is recomputed replacing the mean-field eigenvalues with the QP energies from Equation~\eqref{eq:epsilon_n}.
We enforce eigenvalue self-consistency only in $G$. Inserting the QP energies also in $W_0$ (ev$GW$) reduces the screening, which is not advantageous because the overscreening in $W$ at the PBE level compensates the underscreening due to missing vertex corrections. It has been shown that ev$GW_0$ yields band gaps in good agreement with experiment,\cite{Shishkin2006} while underscreening errors in the ev$GW$ scheme lead to too large band gaps~\cite{Shishkin2006} and overly stretched spectra.\cite{Marom2012b}\par
Higher-level self-consistency schemes, such as fully-selfconsistent $GW$ (sc$GW$)\cite{Caruso2012,Caruso2013} or quasiparticle self-consistent $GW$ (QS$GW$),\cite{Schilfgaarde2006} are expected to restore the 1s QP peak as well, but might not yield better agreement with experiment than ev$GW_0$. It has been shown that sc$GW$ overestimates molecular HOMO excitations\cite{Caruso2016} and band gaps in solids.\cite{Grumet2018} Similar underscreening effects are also expected for core states. 
A first exploratory study seems to confirm this assumption for QS$GW$, reporting an overestimation of 2~eV for 1s core states of small molecules.\cite{Setten2018}\par
ev$GW_0$ is computationally more demanding than a $G_0W_0$ calculation because the QP equation is not only solved for the 1s core states of interest, but repeatedly for all occupied and virtual states until convergence in $G$ is reached. We showed that the core-level QP peak can be also restored in a $G_0W_0$ calculation by using a XC functional with a high fraction of exact exchange as starting point.\cite{Golze2020} We employ the PBEh($\alpha$) functional family with an adjustable amount $\alpha$ of HF exact exchange.\cite{Atalla2013} The XC energy $E_{\mathrm{xc}}$ is given by 
\begin{equation}
 E_{\mathrm{xc}} = \alpha E_x^{\mathrm{EX}} + (1-\alpha)E_{x}^{\mathrm{PBE}} + E_{c}^{\mathrm{PBE}}, \quad \alpha\in[0,1],
\end{equation}
where $E_x^{\mathrm{EX}}$ denotes the HF exchange energy. $E_{x}^{\mathrm{PBE}}$ and $E_{c}^{\mathrm{PBE}}$ are the PBE exchange and correlation energy, respectively. 

In this work, we followed Ref.~\citenum{Golze2020} and used both ev$GW_0$ and $G_0W_0$@PBEh($\alpha$). We analyze how the relativistic corrections we devised affect the two schemes.

\subsection{Relativistic methods}
It has long been recognized that relativistic effects play a large role in the chemistry of heavy elements.\cite{Pyykko1988, Schwerdtfeger2002} In this work, we treat the core states of light elements carbon through fluorine, whose BEs are in the range of 250 - 700~eV, and for which relativistic effects are usually smaller than 1~eV. However, the accuracy required to resolve XPS spectra of 1s excitations of 2nd row elements is in the range of some tenths of an electronvolt and therefore on the same order of magnitude as the relativistic effects.\par
Our relativistic correction scheme for $GW$ is based on two different relativistic KS-DFT methods. The first is the 4-component Dirac Kohn-Sham (4c-DKS) approach, further also referred to as fully-relativistic scheme. The second uses the scalar-relativistic ZORA.\par    
%
\subsubsection{Fully relativistic Dirac approach}
The relativistic description of a non-interacting electron in an external potential $V$ is given by the Dirac equation\cite{Dirac1928} 
\begin{equation}
h_D \Psi = \epsilon \Psi
\label{eq:Dirac_free}
\end{equation}
where $\Psi=\begin{pmatrix}\phi\\\chi\end{pmatrix}$ is a 4-component Dirac spinor, which is comprised of a large component $\phi$ and a small component $\chi$, each of which have two components for the spin functions. The Dirac Hamiltonian $h_D$ is given by 
\begin{equation}
h_D = \begin{pmatrix}
 V & c\bm{\sigma\cdot p} \\
c\bm{\sigma\cdot p} &  -2c^2+V
\end{pmatrix}
\label{eq:dirac_ham}
\end{equation}
where $c$ the speed of light, $\bm{\sigma}$ is a vector of Pauli spin matrices and $\bm{p}$ is the momentum operator. For electron-like states in the non-relativistic limit ($c\rightarrow \infty$), the large component $\phi$ reduces to the wave function of the Schr\"odinger equation, while the small component vanishes. \par
For the case with $N$ interacting electrons, the electronic relativistic Hamiltonian is
\begin{equation}
H_D = \sum_i^N h_D(i) + \sum_{i<j}^N g(i,j),
\label{eq:Dirac}
\end{equation}
where $g(i,j)$ is the electron-electron interaction. In the non-relativistic case, $g(i,j)$ corresponds to the Coulomb operator, where the interaction between two electrons is instantaneous. When including relativity, this cannot be correct because the Coulomb interaction between electrons involves the exchange of photons traveling at the speed of light. The relativistic electron-electron operator is much more complicated than the non-relativistic one and cannot be written in closed form. Its perturbation expansion in terms of $1/c^2$ yields
\begin{align}
g(i,j)  &= \sum_{n=0}^\infty\begin{pmatrix}
\frac{1}{c^2}
\end{pmatrix}^ng_n(i,j) \nonumber \\ 
&= \frac{1}{|\bm{r}_i - \bm{r}_j|} + \mathcal{O}(c^{-2})
\end{align}
where $g_0$ is the instantaneous Coulomb interaction and the 
first order correction $g_1$ is the Breit term,\cite{Breit1929,Kutzelnigg2002} which introduces magnetic and retardation effects. In our 4c-DKS calculations, we include only the $g_0$ term. The contributions from the Breit term are believed to be small\cite{Visscher1996a,Visscher1996b,Quiney1998,Fossgaard2003} and they are therefore neglected in most relativistic calculations.\par
In relativistic KS-DFT,\cite{Rajagopal1973,Rajagopal1978,MacDonald1979,Saue2002,Reiher2009} the XC functional should in principle also include relativistic effects and should be formulated in terms of the four-current density.\cite{Engel2002} The latter is  the basic density variable in relativistic KS-DFT. A relativistic generalization of the local density approximation (RLDA)\cite{MacDonald1979} has been proposed, as well as a semi-empirical gradient corrected variant (RGGA).\cite{Ramana1981,Rajagopal1978} Common practice, however, is to use a non-relativistic XC functional in conjunction with the Dirac kinetic energy,\cite{Engel2002,Fossgaard2003} which is the procedure we follow in this work. We use here a 4c-DKS approach with non-relativistic GGA and hybrid GGA functionals.\par


\subsubsection{Scalar relativistic ZORA approach}
The computational cost for a fully relativistic 4c-DKS approach is significantly higher than for the non-relativistic Schr\"odinger Kohn-Sham (SKS). The scalar relativistic ZORA approximation retains the computational effort of an SKS calculation, and has been shown to capture relativistic effects in good agreement with other scalar-relativistic all-electron schemes.\cite{Lejaeghere2016}\par
The ZORA scheme is derived by solving one of the two coupled equations in Equation~\eqref{eq:Dirac_free} for the small component $\chi$ and inserting it into the other equation, which yields the following (still exact) expression for the unnormalized large component 
\begin{equation}
    \bm{\sigma}\cdot\bm{p}\frac{c^2}{2c^2-V}\bigg(1 + \frac{\epsilon_n}{2c^ 2 - V}\bigg)^{-1}\bm{\sigma}\cdot\bm{p}\phi_n + V\phi_n=\epsilon_n\phi_n.
\label{eq:H_ESC}
\end{equation}
Expanding the parenthetical term as a geometric series yields the regular approximation.\cite{Reiher2009} Retaining only the scalar part of the zeroth order term transforms Equation~\eqref{eq:H_ESC} to
\begin{equation}
    (T^{\mathrm{ZORA}} + V)\phi_n =\epsilon_n\phi_n
    \label{eq:H_ZORA}
\end{equation}
where the ZORA kinetic energy Hamiltonian $T^{\mathrm{ZORA}}$ is defined as\cite{vanLenthe1993}
\begin{equation}
T^{\mathrm{ZORA}} = \bm{p}\cdot\frac{c^ 2}{2c^2-V}\bm{p}.
\label{eq:T_zora}
\end{equation}
Since the potential enters non-linearly in the denominator of Equation~\eqref{eq:T_zora} it is clear that ZORA is gauge dependent, i.e., a constant shift of the electrostatic potential does not lead to a constant shift in the energy. \par
Different methods have been proposed to restore gauge-invariance. One of them is the popular scaled ZORA approximation,\cite{vanLenthe1994} where the eigenvalues are rescaled after self-consistency is reached. Scaled ZORA restores almost, but not completely gauge-invariance. Full gauge-invariance is achieved in the atomic ZORA scheme (aZORA), which we use in this work. In aZORA the potential in the denominator of Equation~\eqref{eq:T_zora} is replaced with the onsite free-atom potential $v_{at}(j)$ near the nucleus, on which the localized basis function $\varphi_j$ is centered. The aZORA Hamiltonian depends therefore explicitly on the atom index of the basis function $\varphi_j$ it acts upon.  We employ the aZORA approach as defined in Refs.~\citenum{Blum2009} and \citenum{Huhn2017} and benchmarked in Ref.~\citenum{Lejaeghere2016}. Since the kinetic term $T^{\mathrm{aZORA}}$ depends on $\varphi_j$, the matrix elements need to be symmetrized to restore Hermiticity, which finally gives 
\begin{align}
    \bra{\varphi_i}T^{\mathrm{aZORA}}\ket{\varphi_j} &= \frac{1}{2}\bra{\varphi_i}\bm{p}\cdot\frac{c^2}{2c^2-v_{at}(j)}\bm{p}\ket{\varphi_j}\nonumber\\ &+ \frac{1}{2}\bra{\varphi_j}\bm{p}\cdot\frac{c^2}{2c^2-v_{at}(i)}\bm{p}\ket{\varphi_i}
\end{align}

While the absolute values of the scaled ZORA eigenvalues are closer to the 4c-DKS reference, \cite{vanLenthe1994} we expect the relative shifts with respect to 4c-DKS, which are relevant for the proposed correction scheme, to be more consistent with aZORA. The reason is that the latter, unlike scaled ZORA, restores the gauge invariance completely. 

\subsection{Atomic relativistic corrections for $\boldsymbol{GW}$}
\label{sec:atomic_correction}
For $GW$, we have developed three simple correction schemes to account for relativistic effects: I) Atomic relativistic corrections are added to the QP energies. II) The aZORA Hamiltonian is used for the underlying DFT calculation and the obtained KS eigenvalues and MOs are used as a starting point for $GW$. III) aZORA is used as in II \textit{and} atomic relativistic corrections are added to the QP energies. The atomic corrections are always added as a post-processing step to the converged QP energies and have been obtained as follows.\par
For scheme I, the atomic relativistic corrections $\Delta\epsilon_{\mathrm{1s,at}}^{\mathrm{SKS}}$ are computed as difference between the non-relativistic SKS 1s eigenvalues $(\epsilon_{\mathrm{1s,at}}^{\mathrm{SKS}})$ and the fully relativistic 4c-DKS 1s eigenvalues $(\epsilon_{\mathrm{1s,at}}^{\mathrm{4c\mbox{-}DKS}})$,
\begin{equation}
\Delta\epsilon_{\mathrm{1s,at}}^{\mathrm{SKS}} = \epsilon_{\mathrm{1s,at}}^{\mathrm{4c\mbox{-}DKS}} - \epsilon_{\mathrm{1s,at}}^{\mathrm{SKS}}.
\label{eq:De1s_at_S}
\end{equation}
The label ``at" indicates that the calculations are performed for a free neutral atom. For scheme III, we use the atomic corrections $\Delta\epsilon_{\mathrm{1s,at}}^{\mathrm{aZORA}}$,
\begin{equation}
\Delta\epsilon_{\mathrm{1s,at}}^{\mathrm{aZORA}} = \epsilon_{\mathrm{1s,at}}^{\mathrm{4c\mbox{-}DKS}} - \epsilon_{\mathrm{1s,at}}^{\mathrm{aZORA}},
\label{eq:De1s_at_Z}
\end{equation}
evaluating the difference to the aZORA 1s eigenvalues ($\epsilon_{\mathrm{1s,at}}^{\mathrm{aZORA}}$) instead to the SKS eigenvalues. The atomic 1s eigenvalues $\epsilon_{\mathrm{1s,at}}^{\mathrm{SKS}}$,
$\epsilon_{\mathrm{1s,at}}^{\mathrm{aZORA}}$ and  $\epsilon_{\mathrm{1s,at}}^{\mathrm{4c\mbox{-}DKS}}$
are computed self-consistently at the PBE level by solving the radial SKS, aZORA and 4c-DKS equations respectively.

\section{\label{sec:methods} Computational details}
 All $GW$ and $\Delta$SCF calculations are performed with the all-electron FHI-aims program package,\cite{Blum2009,Havu2009,Ren2012} which is based on numerically tabulated atom-centered orbitals (NAOs). Core-level BEs from $G_0W_0$, ev$GW_0$ and $\Delta$SCF calculations are calculated for the CORE65 benchmark set introduced in Ref.~\citenum{Golze2020}, which contains 65 1s binding energies of second-row elements (C, N, O and F) for small organic and inorganic molecules. The settings for $G_0W_0$, ev$GW_0$ and $\Delta$SCF are the same as in our previous work\cite{Golze2020} and are summarized in the following.\par
The $\Delta$SCF calculations are performed with the PBE0\cite{Adamo1999,Ernzerhof1999} hybrid functional employing def2 quadruple-$\zeta$ valence plus polarization (def2-QZVP)\cite{Weigend2005} basis sets. The all-electron def2-QZVP Gaussian basis sets are treated numerically in FHI-aims for compliance with the NAO scheme. We decontract the def2-QZVP basis sets to enable a full relaxation of the other electrons in the presence of a core-hole; see Ref.~\citenum{Golze2020} for further details and an explanation of the basis-set choice.\par
For the $GW$ calculations, the QP equation  (Equation~\eqref{eq:epsilon_n}) is always solved iteratively. For the partially self-consistent ev$GW_0$ scheme, we iterate the eigenvalues additionally in $G$. We use the PBE functional\cite{Perdew1996a} as a starting point for the ev$GW_0$ calculations. For $G_0W_0$, we employ the PBEh($\alpha$) hybrid functionals\cite{Atalla2013} for the underlying DFT calculation, where $\alpha$ indicates the fraction of HF exchange in the functional. The core-level BEs are extrapolated to the complete basis set limit to account for the slow convergence of the $GW$ QP energies with respect to basis set size.\cite{Golze2019,Setten2015,Wilhelm2016,Bruneval2012,Bruneval2013} The extrapolation is performed by a linear regression with respect to the inverse of the total number of basis functions using the Dunning basis set family cc-pV$n$Z ($n$=3-6).\cite{Dunning1989,Wilson1996} Details are given in the Supporting Information (SI) in Table~S3 and comprehensive convergence studies are presented in Figure~S1. Furthermore, we use the CD technique\cite{Golze2018} to compute the $GW$ self-energy. The integral over the imaginary frequency axis in Equation~\eqref{eq:cd} is computed using modified Gauss-Legendre grids\cite{Ren2012} with 200 grid points.\par
Relativistic effects for $GW$ are included in three different ways as described in Section~\ref{sec:atomic_correction}. For $\Delta$SCF, we account for relativistic effects self-consistently using the aZORA approximation.\cite{Blum2009} We also apply the atomic relativistic schemes introduced in Section~\ref{sec:atomic_correction} to $\Delta$SCF for comparison. To obtain the atomic relativistic corrections for equations \eqref{eq:De1s_at_S} and \eqref{eq:De1s_at_Z}, the radial DKS, SKS and aZORA-KS equations are solved self-consistently on numerical real-space grids with the DFTATOM code\cite{Certik2013} incorporated in FHI-aims.\par
We investigate the dependence of the relativistic eigenvalue corrections on the molecular environment, XC functional and basis set using the DIRAC program,\cite{DIRAC18, Saue2020} which features a 4c-DKS DFT implementation for the 3D electronic wave function, enabling also molecular calculations. Similar to Equation~\eqref{eq:De1s_at_S}, we define the molecular corrections as
\begin{equation}
\Delta\epsilon_{1s,\mathrm{mol}}^{\mathrm{SKS}} = \epsilon_{1s,\mathrm{mol}}^{\mathrm{4c\mbox{-}DKS}} - \epsilon_{1s,\mathrm{mol}}^{\mathrm{SKS}}, 
\label{eq:De1s_mol}
\end{equation}
where $\epsilon_{1s,\mathrm{mol}}^{\mathrm{4c\mbox{-}DKS}}$ are molecular 1s eigenvalues of the 4c-DKS Hamiltonian. The corresponding non-relativistic eigenvalues $\epsilon_{1s,\mathrm{mol}}^{\mathrm{SKS}}$ are here obtained from a 4c-DKS calculation, resetting the speed of light to the non-relativistic limit ($c\rightarrow\infty$). 

The DIRAC calculations are performed for the molecular structures of the CORE65 benchmark set, excluding the spin-polarized \ce{O2} case, using all-electron Dyall basis sets \cite{Dyall2016} of triple-zeta quality and the PBE functional. We define the difference $\Delta$MOL between molecular and atomic eigenvalue correction as
\begin{equation}
    \Delta\mathrm{MOL} = \Delta\epsilon_{1s,\mathrm{at}}^{\mathrm{SKS}}  - \Delta\epsilon_{1s,\mathrm{mol}}^{\mathrm{SKS}}.
    \label{eq:deltamol}
\end{equation}
The functional dependence of the atomic corrections is assessed for the PBEh($\alpha$) hybrid family. We also study the basis set dependence for the Dyall series\cite{Dyall2016} with reference to the fully converged radial solution from DFTATOM
\begin{equation}
    \Delta\mathrm{BAS} = \Delta\epsilon_{1s,\mathrm{at}}^{\mathrm{SKS}} (\mathrm{dyall}) - \Delta\epsilon_{1s,\mathrm{at}}^{\mathrm{SKS}}(\mathrm{radial}).
    \label{eq:deltabas}
\end{equation}

In pursuit of open materials science, \cite{Himanen/Geurts/Foster/Rinke:2019} we made the results of all relevant calculations available on the Novel Materials Discovery (NOMAD) repository.\cite{NOMAD}

\section{\label{sec:results} Results and discussion}

We first present the atomic relativistic corrections and discuss their dependence on technical and convergence parameter, the XC functional and the molecular environment. We proceed with a discussion of non-relativistic results for the CORE65 benchmark set and demonstrate how our simple correction schemes, based on these atomic corrections, improve the agreement of the computed absolute 1s BEs to experiment.

\subsection{Atomic relativistic corrections}

\begin{figure}
\includegraphics[width=0.95\linewidth]{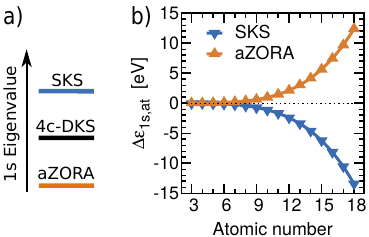}
\caption{a) Schematic of an energy level diagram of 1s eigenvalues comparing non-relativistic Schrödinger Kohn-Sham (SKS), fully relativistic four-component Dirac-Kohn-Sham (4c-DKS) and scalar-relativistic aZORA. b)  Atomic relativistic $\Delta\epsilon_{\mathrm{1s,at}}$ for the spherical SKS (Equation~\eqref{eq:De1s_at_S}) and aZORA Hamiltonian (Equation~\eqref{eq:De1s_at_Z}) with respect to the atomic number.}
\label{fig:De1s}
\end{figure}

\begin{table}
 \caption{\label{tab:atomic} Atomic eigenvalue corrections $\Delta\epsilon_{\mathrm{1s,at}}$ for the SKS (Equation~\eqref {eq:De1s_at_S}) and aZORA Hamiltonian (Equation~\eqref{eq:De1s_at_Z}) for the 2nd and 3rd periods elements. $Z$ indicates the atomic number.}
  \begin{tabular*}{0.99\linewidth}{@{\extracolsep{\fill}}cccc}\toprule
$Z$ & Excitation  & SKS [eV] & aZORA [eV] \\\hline
\hline
3  & Li1s & -0.004245 & 0.003683\\
4  & Be1s & -0.01654  & 0.01516\\
5  & B1s  & -0.05018  & 0.04354\\
6  & C1s  & -0.1176   & 0.1001\\
7  & N1s  & -0.2355   & 0.1996\\
8  & O1s  & -0.4244   & 0.3593\\
9  & F1s  & -0.7080   & 0.6000\\
10 & Ne1s &  -1.113   & 0.9456\\
11 & Na1s & -1 .658   & 1.4423\\
12 & Mg1s & -2.387    & 2.1164\\
13 & Al1s & -3.360    & 3.0092\\
14 & Si1s & -4.607    & 4.1613\\
15 & P1s  & -6.180    & 5.6199\\
16 & S1s  & -8.129    & 7.4357\\
17 & Cl1s & -10.51    & 9.6644\\
18 & Ar1s & -13.39    & 12.365\\\bottomrule
  \end{tabular*}
\end{table}

Figure~\ref{fig:De1s}(a) shows a sketch of the 1s eigenvalues from the non-relativistic SKS, 4c-DKS, and scalar relativistic aZORA calculations. The SKS eigenvalues are generally overestimated with respect to the 4c-DKS reference, while aZORA underestimates the 1s eigenvalues by nearly as much. The atomic eigenvalue corrections $\Delta\epsilon_{\mathrm{1s,at}}$ for SKS (Equation~\eqref{eq:De1s_at_S}) and aZORA (Equation~\eqref{eq:De1s_at_S}) are given in Table~\ref{tab:atomic}. The SKS corrections are negative and increase in magnitude with atomic number, ranging from $-4$~meV for Li to $-13.4$~eV for Ar. The aZORA corrections are positive and increase from 4~meV (Li) to 12.4~eV (Ar). 

The atomic corrections given in Table~\ref{tab:atomic} are visualized in Figure~\ref{fig:De1s}(b). We observe that the magnitude of the atomic corrections for both SKS and aZORA, depends on the fourth power of atomic number $Z$. This dependence is known from the relativistic correction to the exact energy of the hydrogenic orbital, whose leading order term in a perturbative expansion scales as the fourth power of $Z$. \cite{Dirac1928, Gordon1928, Darwin1928, Eides2007} As the 1s orbitals are poorly-screened by the outer orbitals, the magnitude of the relativistic correction trends similarly to that of the unscreened hydrogenic orbitals.\par


\begin{figure}
\includegraphics[width=0.95\linewidth]{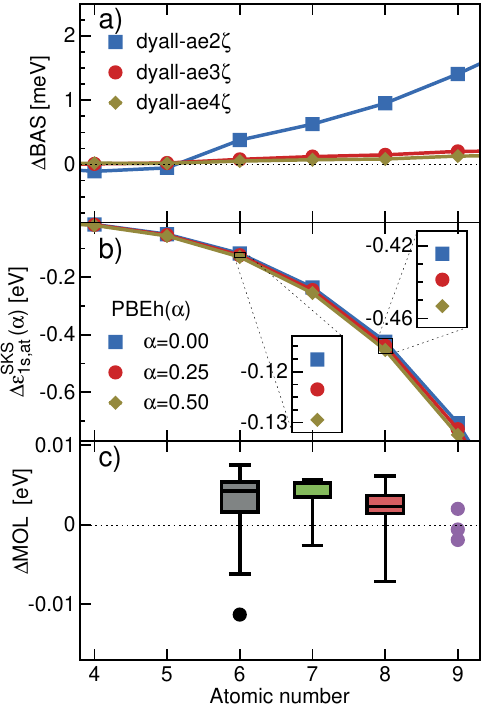}
\caption{Basis set convergence and dependence of the atomic corrections on the XC functional and molecular environment.
a) Difference $\Delta$BAS  as defined in Equation~\eqref{eq:deltabas} between the Dyall all-electron basis set at the double, triple and quadruple-$\zeta$ levels\cite{Dyall2016} and the radial solution on numeric real-space grids.
b) Atomic eigenvalue correction $\Delta\epsilon_{\mathrm{1s,at}}^{\mathrm{SKS}}$ (Equation~\eqref{eq:De1s_at_S}) computed with the PBEh($\alpha$) functional with 3 different values of $\alpha$.
c) Difference $\Delta$MOL as defined in Equation~\eqref{eq:deltamol} between the molecular eigenvalue correction and atomic value for C1s, N1s, O1s and F1s excitations of the CORE65 benchmark set. Bars contain the 2nd and 3rd quartiles, whiskers extend to encompass 95\% of the results, and outliers are shown as dots.
}
\label{fig:dirac_mol}
\end{figure}
The results reported in Table~\ref{tab:atomic} are obtained from the self-consistent solutions of the radial SKS and 4c-DKS equations. The radial equations enforce a spherical symmetry of the solution. However, most atoms have ground states with non-spherical symmetry. For the second-row, this applies to B, C, O and F. For these elements, the spherical solutions are too high in total energy by several tenths of eV and assume fractional occupation numbers. The 1s eigenvalue corrections $\Delta\epsilon_{\mathrm{1s,at}}^{\mathrm{SKS}}$ obtained from the radial SKS and 4c-DKS equation are therefore an approximation. To estimate the error introduced by this approximation, we solved the 3D SKS equations for the free neutral atom and compared the 1s eigenvalues of the spherical and non-spherical solution. The spherical solution is also obtained with the 3D equations and is identical to the radial one, if we do not break the symmetry and enforce integer occupations. The non-spherical 3D solution is obtained by employing occupation constraints. We find that the difference in the absolute 1s eigenvalues between spherical and non-spherical solution is less than 50~meV, see Table~S1 (SI), which is an order of magnitude smaller than the relativistic corrections themselves. The error in the relative values, $\Delta\epsilon_{\mathrm{1s,at}}^{\mathrm{SKS}}$, is expected to be even smaller and we conclude thus that the radial approximation is sufficient. 
\par
The radial calculations are performed on a numeric real-space grid, which can be easily converged, whereas the 3D calculations rely on relativistic all-electron Gaussian basis sets, potentially introducing a basis set incompleteness error. Figure~\ref{fig:dirac_mol}(a) shows the basis set convergence of the Dyall series with respect to the radial solution. At the double-$\zeta$ level, the error is within a few meV, and for the relevant 1s states, we reach convergence already at the triple-$\zeta$ level, see Figure~\ref{fig:dirac_mol}(a). We use the quadruple-$\zeta$ basis set for the calculations shown in Figure~\ref{fig:dirac_mol}(b) and the triple-$\zeta$ basis set for the calculations in Figure~\ref{fig:dirac_mol}(c). \par 


%
In Figure~\ref{fig:dirac_mol}(b), we examine the dependence of the atomic eigenvalue correction $\Delta\epsilon_{\mathrm{1s,at}}^{\mathrm{SKS}}$ on the fraction $\alpha$ of exact HF exchange in the PBEh($\alpha$) functional for $\alpha=0$ (PBE), $\alpha=0.25$ (PBE0) and $\alpha=0.5$. The magnitude of the eigenvalue correction shows a slight dependence on $\alpha$, increasing by an amount that is proportional to the fraction of exact exchange. At first glance, the $\alpha$ dependence seems more pronounced for heavier elements. However, this is only true for the absolute values: Setting the PBE functional as reference and comparing to PBEh($\alpha=0.5$), the magnitude of the  atomic correction increases by 12~meV for carbon~1s, which corresponds to 10.2\%, and by 40~meV for fluorine 1s, which, however, corresponds only to 5.7\%. In fact, the $\alpha$ dependence seems to decrease with the atomic number when comparing relative deviations; see Table~S6 (SI) for the tabulated values. For all elements listed in Table~\ref{tab:atomic}, we find that the $\alpha$ dependence is an order of magnitude smaller than the relativistic correction itself. We thus neglect it when applying our relativistic correction schemes to the 1s QP energies from $G_0W_0@$PBEh.\par
In Figure~\ref{fig:dirac_mol}(c), we compare the atomic eigenvalue correction (Equation~\eqref{eq:De1s_at_S}) to the molecular eigenvalue correction (Equation~\eqref{eq:De1s_mol}). For most of the excitations considered, the atomic eigenvalue correction slightly underestimates the molecular correction, but the difference between the two is under 5 meV for 49 of the 63 excitations considered, with a maximum deviation of 12.6 meV. The distribution of these differences is similar for the core excitations of different elements, and is small enough in comparison with the magnitude of the eigenvalue correction to justify the use of the atomic values irrespective of the chemical environment.\par
The atomic SKS corrections reported in Table~\ref{tab:atomic} are very similar to the atomic corrections published in Ref. \citenum{Bellafont2016b} for second-row elements B$\rightarrow$F. The atomic corrections in Ref.~\citenum{Bellafont2016b} are 10-40~meV larger than ours and were computed by comparing four-component Dirac-Hartree-Fock (DHF) energies with non-relativistic HF energies. Our analysis of different PBEh functionals in Figure~\ref{fig:dirac_mol}(b) suggest that these differences must be partly attributed to the exchange treatment. The remaining differences might be due to usage of non-relativistic basis sets in combination with the 4c-DHF Hamiltonian in Ref.~\citenum{Bellafont2016b}. The atomic SKS corrections are also surprisingly similar to the corrections derived for second-period elements in an early work from the 1960s based on Pauli perturbation theory of charged 2-electron atoms.\cite{Pekeris1958,Mukherjee1985} Pauli perturbation theory is based on the first order in the expansion of Equation~\eqref{eq:H_ESC} in terms of $1/c^2$. It is highly singular in the deep-core region\cite{Chang1986} and has been largely replaced by the regular approximation, which expands Equation~\eqref{eq:H_ESC} in terms of  $\frac{\epsilon}{2c^2 - V}$. The correspondence worsens when valence electrons are included. \cite{Triguero1999} \par

\begin{figure}
\includegraphics[width=0.95\linewidth]{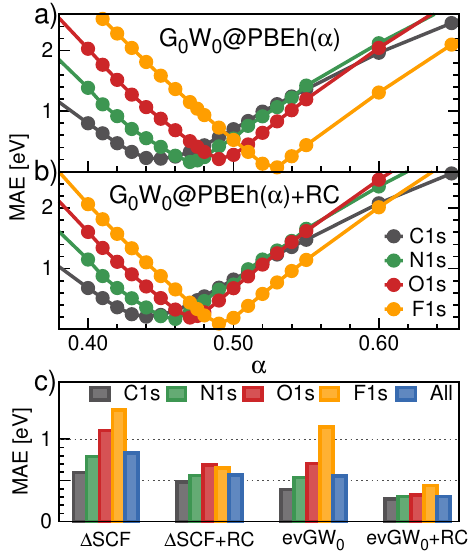}
\caption{Mean absolute error (MAE) of absolute 1s BEs with respect to experiment for the CORE65 benchmark set. MAE for $G_0W_0$@PBEh dependent on the fraction of exact exchange $\alpha$ in the PBEh functional (a) without and (b) with atomic relativistic correction (RC). (c) MAE for $\Delta$SCF and ev$GW_0$@PBE, without and with RC.}

\label{fig:alpha}
\end{figure}

\subsection{Non-relativistic quasiparticle energies}

In our previous work\cite{Golze2020} we briefly discussed the effect of relativistic corrections, comparing non-relativistic and relativistic 1s BEs from ev$GW_0$ to experiment. We will now analyze the non-relativistic ev$GW_0$ calculations in more detail and additionally include the non-relativistic $G_0W_0@$PBEh and $\Delta$SCF results in the discussion.\par
\begin{figure}
\includegraphics[width=0.95\linewidth]{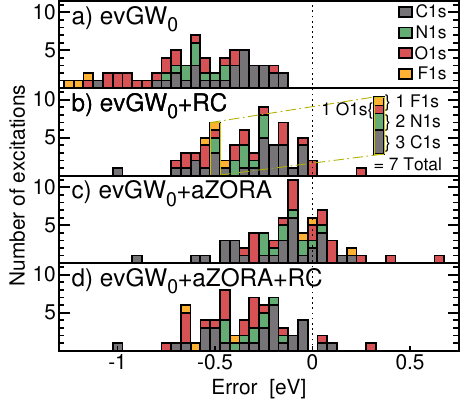}
\caption{
Distribution of errors with respect to experiment for absolute 1s BEs of the CORE65 benchmark set, where Error$_i$= $\text{BE}_i^{\text{theory}}-\text{BE}_i^{\text{exp}}$. Note that the histogram is stacked. Three different relativistic correction (RC) schemes are compared. (a) Non-relativistic ev$GW_0$, (b) ev$GW_0$ adding the atomic corrections $\epsilon_{\mathrm{1s,at}}^{\mathrm{SKS}}$ to the QP energies, (b) ev$GW_0$ using aZORA in the underlying DFT calculation, (c) ev$GW_0$ using aZORA as in (b) and adding atomic corrections $\epsilon_{\mathrm{1s,at}}^{\mathrm{aZORA}}$ to the QP energies.}
\label{fig:err_hist}
\end{figure}

Figure~\ref{fig:alpha} displays the mean absolute error (MAE) of the absolute 1s BEs with respect to experiment for the CORE65 benchmark set. 
The MAEs obtained from non-relativistic ev$GW_0$ calculations increase with atomic number (Figure~\ref{fig:alpha}(c)) and the magnitude of this increase is within the range of the atomic relativistic corrections given in Table~\ref{tab:atomic}. The distribution of these errors is shown in Figure~\ref{fig:err_hist}(a), where the grouping in species is evident. ev$GW_0$ systematically underestimates the 1s BEs for all 65 excitations. The non-relativistic $\Delta$SCF calculations underestimate the 1s BEs as well and the MAEs show a very similar trend with respect to atomic number, see Figure~\ref{fig:alpha}(c). Comparing the overall MAE for the non-relativistic calculations, we find that the MAE for $\Delta$SCF is with 0.71~eV slightly larger than the 0.55~eV MAE for ev$GW_0$. \par
Relativistic effects are also apparent when considering the optimal $\alpha$ for use in a $G_0W_0$@PBEh($\alpha$) scheme. Figure~\ref{fig:alpha}(a) shows the MAE for non-relativistic $G_0W_0$@PBEh($\alpha$) calculations with respect to the fraction of exact exchange $\alpha$. These calculations have been carried for a subset of 43 excitations of the CORE65 benchmark set, for which the mapping between core state and atom does not require analysis of, e.g., MO coefficients. In our previous work\cite{Golze2020}, we reported the $\alpha$ dependence of the MAE including relativistic effects (Figure~\ref{fig:alpha}(b)) and found that the smallest MAE is obtained for $\alpha$ values around 0.45. For MAEs smaller than 0.45, the BEs are underestimated and for larger $\alpha$ values increasingly overestimated. For the non-relativistic results we observe a much stronger species dependence of the optimal $\alpha$ value. As the non-relativistic Hamiltonian underestimates the core-level BE, increasing the exact exchange reduces the screening, resulting in a larger BE. An increase in $\alpha$ can thus offset the relativistic error. Comparing the MAE from non-relativistic $G_0W_0$@PBEh($\alpha$) calculations (Figure \ref{fig:alpha}(a)), we find that the optimal $\alpha$ indeed increases with atomic number, from 0.44 for C1s excitations to about 0.55 for F1s.

\subsection{Atomic and scalar-relativistic correction schemes}
We investigate three simple schemes to account for relativistic corrections (RC) in 1s core-level BEs from $GW$. Additionally, we discuss the application of these three schemes to $\Delta$SCF. 
The first approach is to add the atomic corrections  $\Delta\epsilon_{\mathrm{1s,at}}^{\mathrm{SKS}}$ defined in Equation~\eqref{eq:De1s_at_S} to the QP energies and corresponds to the scheme we employed in our latest work.\cite{Golze2020} We label scheme I ``method+RC". The second is to use aZORA for the underlying DFT calculations, and use the aZORA eigenvalues and MOs as the starting point for the $GW$ calculation. We refer to scheme II as ``method + aZORA''. In the third scheme, we use scheme II to obtain the QP energies and add the atomic corrections $\Delta\epsilon_{\mathrm{1s,at}}^{\mathrm{aZORA}}$ defined in Equation~\eqref{eq:De1s_at_Z} afterwards. We label scheme III ``method + aZORA + RC".\par
For ev$GW_0$ we explored also a variant of scheme I, where we added the atomic corrections to the DFT eigenvalues instead to the QP energies. These corrected eigenvalues were then used as starting point for the $evGW_0$ calculation. This pre-correction variant yields with a mean absolute difference of 16~meV BEs that are extremely similar to the ones from ev$GW_0$+RC. Adding the atomic correction as post-processing step is transferable to non-relativistic $GW$ results obtained from any code and we thus disregard the pre-correction variant in the following.
\par
Compared with the non-relativistic energies, the ev$GW_0$+RC scheme reduces the error with respect to experiment, as shown in Figure~\ref{fig:err_hist}(b). The errors are more tightly distributed, and the clustering by species is no longer evident. Generally, the BEs are still underestimated. However, the overall MAE is reduced from 0.55  to 0.3~eV and is now well within the accuracy required for chemical analysis. Furthermore, the species-dependence in the MAE is largely eliminated; see Figure~\ref{fig:alpha}(c) and Table~\ref{tab:MAE}. Solely the MAE for the F1s excitations is with 0.44~eV slightly larger than for the other elements. This might be attributable to poor statistics since our benchmark set contains only 3 F1s excitation. \par
Scheme~I has also been successfully employed for $G_0W_0@$PBEh. The range of optimal $\alpha$ is reduced by a factor of two for the  $G_0W_0@$PBEh($\alpha$)+RC scheme vis-a-vis the non-relativistic one, see Figure \ref{fig:alpha}(a,b). With the relativistic correction, the value of $\alpha$ that minimizes the MAE ranges from 0.44 for C1s excitations to 0.49 for F1s excitations. This shows also in a slight species-dependent of the MAE value we reported for the $G_0W_0@$PBEh($\alpha$=0.45) results with RC earlier.\cite{Golze2020}

\begin{table*}[t!]
\caption{\label{tab:MAE} Mean absolute error (MAE) and mean error (ME) in eV with respect to experiment, by species and in aggregate for absolute BEs of the CORE65 benchmark set. The error for excitation $i$ is defined as Error$_i$= $\text{BE}_i^{\text{theory}}-\text{BE}_i^{\text{exp}}$.}
\begin{tabular*}{0.99\linewidth}{p{0.8cm}KKKKKKKKKKKKKKKKK}\toprule
\multirow{2}{*}{\parbox{0.7cm}{\centering core-level}} 
& \multicolumn{2}{c}{\parbox{1.84cm}{$\Delta$SCF}} 
& \multicolumn{2}{c}{\parbox{1.84cm}{$\Delta$SCF+RC}} 
& \multicolumn{2}{c}{\parbox{1.84cm}{$\Delta$SCF+aZORA}} 
& \multicolumn{2}{c}{\parbox{1.84cm}{$\Delta$SCF+ aZORA+RC}} 
& \multicolumn{2}{c}{ev$GW_0$}  
& \multicolumn{2}{c}{ev$GW_0$+RC}     
& \multicolumn{2}{c}{ev$GW_0$+aZORA}     
& \multicolumn{2}{c}{\parbox{1.84cm}{ev$GW_0$+ aZORA+RC}}   \\

\cmidrule(l{0.5em}r{0.5em}){2-3}\cmidrule(l{0.5em}r{0.5em}){4-5}\cmidrule(l{0.5em}r{0.5em}){6-7}\cmidrule(l{0.5em}r{0.5em}){8-9}\cmidrule(l{0.5em}r{0.5em}){10-11}\cmidrule(l{0.5em}r{0.5em}){12-13}\cmidrule(l{0.5em}r{0.5em}){14-15}\cmidrule(l{0.5em}r{0.5em}){16-17}
    &   
    \multicolumn{1}{c}{MAE}  & \multicolumn{1}{c}{ME}  &   \multicolumn{1}{c}{MAE}  & \multicolumn{1}{c}{ME} & \multicolumn{1}{c}{MAE}  & \multicolumn{1}{c}{ME} & \multicolumn{1}{c}{MAE}  & \multicolumn{1}{c}{ME} & \multicolumn{1}{c}{MAE}  & \multicolumn{1}{c}{ME} & \multicolumn{1}{c}{MAE}  & \multicolumn{1}{c}{ME} & \multicolumn{1}{c}{MAE}  & \multicolumn{1}{c}{ME} & \multicolumn{1}{c}{MAE}  & \multicolumn{1}{c}{ME} \\\hline
all & 0.83 & -0.83 & 0.57 & -0.57 & 0.33 & -0.31 & 0.46 & -0.46 & 0.55 & -0.55 & 0.30 &  -0.29  & 0.18  & -0.07 & 0.32 & -0.29\\
C1s & 0.60  & -0.60 & 0.48 & -0.48 & 0.36 & -0.36 & 0.52 & -0.52 & 0.39 & -0.39 & 0.27 &  -0.27  & 0.21  & -0.15 & 0.27 & -0.25\\
N1s & 0.79 & -0.79 & 0.56 & -0.56 & 0.32 & -0.32 & 0.52 & -0.52 & 0.54 & -0.54 & 0.30 &  -0.30  & 0.10  & -0.07 & 0.27 & -0.27\\
O1s & 1.11 & -1.11 & 0.69 & -0.68 & 0.32 & -0.27 & 0.64 & -0.63 & 0.71 & -0.71 & 0.32 &  -0.28  & 0.19  &  0.02 & 0.38 & -0.34\\
F1s & 1.36 & -1.36 & 0.65 & -0.65 & 0.12 &  0.03 & 0.57 & -0.57 & 1.15 & -1.15 & 0.44 &  -0.44  & 0.10  &  0.08 & 0.52 & -0.52\\ \bottomrule
\end{tabular*}
\end{table*}

\begin{figure}
\includegraphics[width=0.95\linewidth]{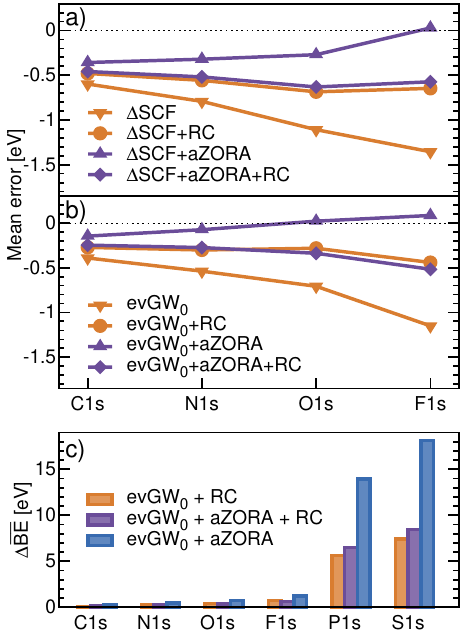}
\caption{(a) and (b): Mean error with respect to experiment for 1s BEs from (a) $\Delta$SCF and (b) ev$GW_0$ for the CORE65 benchmark set, where the error for excitation $i$ is defined as Error$_i$= $\text{BE}_i^{\text{theory}}-\text{BE}_i^{\text{exp}}$. Three relativistic schemes are compared to the non-relativistic results. (c) Average size of the relativistic correction for 1s BEs from ev$GW_0$ for the CORE65 benchmark set with five additional molecules containing third period elements, where $\Delta \text{BE}=\text{BE}_{\text{relativistic}}-\text{BE}_{\text{non-relativistic}}$. For ev$GW_0$+RC, $\Delta$BE corresponds to the negative of the atomic corrections $\Delta\epsilon_{\mathrm{1s,at}}^{\mathrm{SKS}}$  given in Table~\ref{tab:atomic}.}
\label{fig:me}
\end{figure}

Judging by the MAE alone (Table~\ref{tab:MAE}), the ev$GW_0$+aZORA results are an improvement over the ev$GW_0$+RC scheme. The overall MAE is 0.18~eV. Their distribution (Figure~\ref{fig:err_hist}(c)) is more centered. A slight clustering by species is observed, although this is not as obvious as for the non-relativistic results shown in Figure~\ref{fig:err_hist}(a). In contrast to the non-relativistic values, the 1s excitations of the lighter elements, such as carbon, tend to be more underestimated than the 1s BEs of oxygen.\par
The ev$GW_0$+aZORA+RC scheme performs worse than ev$GW_0$+aZORA approach, with an error distribution similar to ev$GW_0$+RC, see Figure \ref{fig:err_hist}(d). The MAE for the individual species are in the same range as for ev$GW_0$+RC, as is the overall MAE with 0.32~eV, see Table~\ref{tab:MAE}. The ev$GW_0$+aZORA+RC scheme is the most sophisticated among the three relativistic corrections discussed here: scalar-relativistic effects are included in the MOs used as starting point for the ev$GW_0$ calculation and the QP energies are corrected with respect to the fully-relativistic atomic reference. It is thus surprising that it performs worse than ev$GW_0$+aZORA. The similar performance of ev$GW_0$+RC and ev$GW_0$+aZORA+RC rather implies that the effect of including relativistic effects in the MOs is minimal.\par
To further investigate this surprising behavior, we visualized the mean errors (MEs) for C1s, N1s, O1s and F1s in Figure~\ref{fig:me}(b). The error is defined as $\text{BE}_{\mathrm{theory}}-\text{BE}_{\mathrm{experiment}}$. Negative MEs indicate thus a systematic underestimation of the BEs with respect to experiment. For the non-relativistic results, the absolute value of the ME corresponds directly to the MAE and the ME is increasingly negative with atomic number. The MEs for ev$GW_0$+RC and ev$GW_0$+aZORA+RC are negative and show almost no species dependence, which is in agreement with our previous analysis of the MAE and the error distribution in
Figure~\ref{fig:err_hist}. For ev$GW_0$+aZORA, however, we observe a trend that is reverse to the non-relativistic results. The ME increases with atomic number and becomes even positive for F1s.\par
Comparing non-relativistic with the relativistic BEs, we find that the size of the relativistic correction is 2-3 times larger with ev$GW_0$+aZORA than with the other two schemes, see Figure~\ref{fig:me}(c). In combination with the upwards trend observed for the ME, this implies that aZORA is overestimating the relativistic correction for 1s states. This reflects the well-known tendency of aZORA to overestimate the relativistic correction to core-state eigenvalues.\cite{vanLeeuwen1994,Sundholm2002} For the 2nd period elements, where the relativistic error ranges from $0.2-0.7$~eV, the aZORA overcorrection compensates in part a chronic underestimation of the BEs. While this error cancellation may seem fortuitous for 2nd row elements, the rapid growth of the relativistic correction with the atomic number implies that ev$GW_0$+aZORA might lead to large errors for 1s BEs of heavier elements. To illustrate this, we analyze the relativistic corrections for a small number of phosphorus and sulfur containing small molecules alongside those for the CORE65 benchmark set: \ce{H2S}, \ce{SO2}, \ce{PH3}, \ce{PF3} and \ce{PF5} (see Figure~\ref{fig:me}(c)). The relativistic corrections obtained with the ev$GW_0$+aZORA scheme are more than 10~eV larger than with ev$GW_0$+RC and ev$GW_0$+aZORA+RC. Also the difference between the ev$GW_0$+aZORA+RC and ev$GW_0$+RC, which is negligible for 2nd period elements, becomes more significant: 2.1~eV for the P1s excitations, and  2.6~eV for the S1s. This suggests that the use of a scalar-relativistic reference for the underlying DFT calculation becomes more relevant as the magnitude of relativistic effects increase. However the effect of the relativistic reference is only about one third the magnitude of the relativistic correction for these states.\par
%
%
We applied the three correction schemes also to 1s BEs obtained from $\Delta$SCF and plotted the MEs by species in Figure~\ref{fig:me}(a). We observe the same trends as for ev$GW_0$. Note the similarity between Figure~\ref{fig:me}(a) and (b). $\Delta$SCF+RC or $\Delta$SCF+aZORA+RC largely eliminate the species dependence of the ME. Although it is not as marked as for ev$GW_0$+aZORA, the MEs increase also slightly with atomic number for $\Delta$SCF+aZORA. The size of the relativistic correction is two times larger with $\Delta$SCF+aZORA than with $\Delta$SCF+RC or $\Delta$SCF+aZORA+RC, which suggests that the relativistic 1s corrections are also overestimated at the $\Delta$SCF+aZORA level. $\Delta$SCF entails the difference in total energy between an ionized and neutral systems. However, the ZORA overcorrection of core states propagates to the final calculation of the BE in approximately the same manner as in the 1s levels. This can be seen from Slater transition state theory: \cite{Slater1951,Slater1971} In $\Delta$SCF we calculate the energy difference $\hbar\omega = E(N-1) - E(N)$, where $N$ is the number of electrons and $E(N-1)$ and $E(N)$ are the total energies of the core-ionized and neutral systems, respectively. This energy difference is approximately related to the 1s eigenvalue by $\hbar\omega = \partial E/\partial n_{\mathrm{1s}} + \mathcal{O}((\delta n_{\mathrm{1s}})^3)$ and $\partial E/\partial n_{\mathrm{1s}} = \epsilon_{\mathrm{1s}}$, where $n_{\mathrm{1s}}$ is the occupation number of the 1s state.\cite{Williams1975}  This is an important detail to consider when comparing the performance of XC functionals for $\Delta$SCF calculations of 1s excitations since the relativistic treatment makes a difference. For example, a recent study\cite{Kahk2019} with the SCAN functional uses the $\Delta$SCF approach in combination with scaled ZORA, while an atomic correction scheme was used for a similar benchmark study\cite{Bellafont2016b} with the TPSS functional. Both studies report very good agreement with experiment. However, it is difficult to judge, which functional performs better, since the relativistic effects are not treated on equal footing.

Both schemes that consistently improve the agreement with experiment, ev$GW_0$+RC and ev$GW_0$+aZORA+RC, chronically underestimate the 1s BEs. This might be attributed to the broadening of the experimental spectra due to vibration effects, while $GW$ yields vertical excitation energies. It has been demonstrated for $GW$-computed excitations of frontier orbitals that the deviation to experiment can be reduced to 0.1~eV when fully resolving the vibrational structure based on Franck-Condon multimode analysis, performed as post-processing step to the $GW$ calculation.\cite{Gallandi2015} This level of accuracy is often not required to resolve most XPS spectra, but could be in principle reached by applying the same approach.

\section{Conclusion}
\label{sec:conclusion}
Relativistic corrections for 1s core-level energies from $GW$ have been derived for non-relativistic and scalar-relativistic starting points. We have investigated three schemes for 1s QP energies from ev$GW_0$: A post-$GW$ atomic correction (ev$GW_0$+RC) using a non-relativistic reference, employing an aZORA reference (ev$GW_0$+aZORA), and employing an aZORA reference along with a post-$GW$ atomic correction (ev$GW_0$+aZORA+RC). All three schemes improve agreement with experiment. The ev$GW_0$+RC and ev$GW_0$+RC+aZORA schemes reduce the mean absolute error to about 0.3~eV  and eliminate the species dependence. The ev$GW_0$+aZORA scheme further reduces the overall MAE to 0.2~eV, but does so inconsistently, due to a species-dependent overcorrection. 

The similarity of the results for the ev$GW_0$+RC and ev$GW_0$+aZORA+RC schemes indicates that the use of the scalar-relativistic reference has no significant effect on the result. Of the two, the ev$GW_0$+RC scheme offers the further advantage that it is readily applicable in codes that have not implemented the aZORA Hamiltonian. We have shown that the derived corrections for the non-relativistic reference improve also consistently core-level energies from $G_0W_0$ and $\Delta$SCF, suggesting that they are generally applicable to core-level BEs from different theoretical methods. \par
The ev$GW_0$+RC and ev$GW_0$+aZORA+RC schemes correct the non-relativistic values in a consistent manner, and therefore form a solid foundation for further development and refinement. Further refinements may include the inclusion of vibrational effects to further improve the accuracy, in particular for C1s excitations, which are generally subject to very small chemical shifts. The development of relativistically-corrected $GW$ also paves the way for the accurate calculation of XPS the in condensed phase systems, where $\Delta$-approaches are problematic. Relativistic corrections will also improve the accuracy of X-ray absorption spectra from the Bethe-Salpeter equation, which employs the quantities computed in the $GW$ calculations.\cite{Salpeter1951,Liu2020}

\begin{acknowledgements}
 We thank the CSC - IT Center for Science for providing computational resources. D. Golze acknowledges financial support by the Academy of Finland (Grant 
No. 316168).
\end{acknowledgements}

\section*{Supporting Information Available}

Comparison of 1s eigenvalues for spherical and non-spherical solutions to the KS equations for neutral atoms (Table~S1). Plot of basis set dependence of extrapolation scheme (Figure~S1). Convergence of KS 1s eigenvalues and total energies for cc-pVnZ and NAO-VCC-nZ basis set series (Table~S2). Scalar-relativistic ev$GW_0$@PBE+aZORA results for basis set series cc-pVnZ (n=3,6), extrapolated values standard errors, and correlation coefficients. (Table~S3). Results and experimental data for CORE65 benchmark set for ev$GW_0$+aZORA and (Table~S4) and for $\Delta$SCF+aZORA, and precorrected non-relativistic ev$GW_0$ (Table~S5). Tabulated values of Figure 2(b) (Table~S6). Difference between pre- and post-corrected schemes for the CORE65 Benchmark set (Figure~S2).

\section*{References}
\bibliography{main}

\begin{thebibliography}{134}
\bibitem [{NOM()}]{NOMAD}%
  \BibitemOpen
  \href@noop {} {}\bibinfo {note} {Submission to NOMAD repository will follow
  after inclusion of reviewer's comments upon publication.}\BibitemShut {Stop}%
\end{thebibliography}
\end{document}